\documentclass{article} 
\usepackage{iclr2023_workshop,times}

\usepackage{amsmath,amsfonts,bm}

\def\eqref#1{equation~\ref{#1}}

\def\1{\bm{1}}

\DeclareMathAlphabet{\mathsfit}{\encodingdefault}{\sfdefault}{m}{sl}
\SetMathAlphabet{\mathsfit}{bold}{\encodingdefault}{\sfdefault}{bx}{n}

\newcommand*{\PP}{\text{P}}
\usepackage[backref=page]{hyperref}
\PassOptionsToPackage{linktocpage}{hyperref}
\usepackage{url}
\usepackage{sidecap}

\usepackage{xcolor}
\definecolor{darkpurple}{rgb}{0.5,0,0.5}
\definecolor{mydarkgrey}{rgb}{0.27, 0.27, 0.27}
\usepackage{wrapfig}
\usepackage{marginnote}
\usepackage{overpic}
\usepackage{multicol}
\usepackage[page,header]{appendix}
\usepackage{titletoc}
\definecolor{antiquefuchsia}{rgb}{0.57, 0.36, 0.51}
\definecolor{cadetblue}{rgb}{0., 0.62, 0.63}
\definecolor{brightmaroon}{rgb}{0.76, 0.13, 0.28}
\definecolor{mymaroon}{rgb}{0.62, 0.05, 0.19}
\definecolor{mymaroon2}{cmyk}{0, 0.739, 0.614, 0.368}
\definecolor{darkpurple}{rgb}{0.5,0,0.5}
\definecolor{mydarkgrey}{rgb}{0.27, 0.27, 0.27}
\definecolor{mymagenta}{rgb}{0.8125, 0, 0.8125}

\renewcommand*{\backref}[1]{}
\renewcommand*{\backrefalt}[4]{%
  \ifcase#1
    (not cited)%
  \or
    \textcolor{gray}{\emph{(cited on page:~#2)}}%
  \else
    \textcolor{gray}{\emph{(cited on pages:~#2)}}%
  \fi
}

\usepackage{scalerel}
\usepackage{tikz}
\usetikzlibrary{svg.path}

\definecolor{orcidlogocol}{HTML}{A6CE39}
\tikzset{
  orcidlogo/.pic={
    \fill[orcidlogocol] svg{M256,128c0,70.7-57.3,128-128,128C57.3,256,0,198.7,0,128C0,57.3,57.3,0,128,0C198.7,0,256,57.3,256,128z};
    \fill[white] svg{M86.3,186.2H70.9V79.1h15.4v48.4V186.2z}
                 svg{M108.9,79.1h41.6c39.6,0,57,28.3,57,53.6c0,27.5-21.5,53.6-56.8,53.6h-41.8V79.1z M124.3,172.4h24.5c34.9,0,42.9-26.5,42.9-39.7c0-21.5-13.7-39.7-43.7-39.7h-23.7V172.4z}
                 svg{M88.7,56.8c0,5.5-4.5,10.1-10.1,10.1c-5.6,0-10.1-4.6-10.1-10.1c0-5.6,4.5-10.1,10.1-10.1C84.2,46.7,88.7,51.3,88.7,56.8z};
  }
}

\newcommand\orcidicon[1]{\href{https://orcid.org/#1}{\mbox{\scalerel*{
\begin{tikzpicture}[yscale=-1,transform shape]
\pic{orcidlogo};
\end{tikzpicture}
}{|}}}}

\definecolor{orcidlogocol}{HTML}{A6CE39}

\usepackage{nicefrac}       
\hypersetup{colorlinks,urlcolor=darkpurple, citecolor=mymaroon, linkcolor=magenta}
\usepackage{overpic}
\usepackage{amssymb}
\usepackage{mathrsfs}
\title{\centering Geometric constraints improve inference of sparsely observed stochastic dynamics }

{\centering
\author{Dimitra Maoutsa \thanks{\href{https://dimitra-maoutsa.gitlab.io/}{https://dimitra-maoutsa.gitlab.io/} } \,\orcidicon{0000-0002-3553-8658} \\
Technical University of Berlin\\
Germany\\
\texttt{dimitra.maoutsa@\{tu-berlin.de ; gmail.com\}  } 
}
}
\iclrfinalcopy 
\begin{document}

\maketitle

\begin{abstract}
The dynamics of systems of many degrees of freedom evolving on multiple scales are often modeled in terms of stochastic differential equations. Usually the structural form of these equations is unknown and the only manifestation of the system's dynamics are observations at discrete points in time. Despite their widespread use, accurately inferring these systems from sparse-in-time observations remains challenging. 
Conventional inference methods either focus on the temporal structure of observations, neglecting the geometry of the system's invariant density, or use geometric approximations of the invariant density, which are limited to conservative driving forces. To address these limitations, here, we introduce a novel approach that reconciles these two perspectives. 
We propose a path augmentation scheme that employs data-driven control to account for the geometry of the invariant system's density. 
Non-parametric inference on the augmented paths, enables efficient identification of the underlying deterministic forces of systems observed at low sampling rates.


\end{abstract}
\section{Introduction}
Unraveling a system's governing equations from time-series observations is often crucial for understanding unexplained natural phenomena. The goal is to find a mathematical representation that aligns with observational data and provides a comprehensive phenomenological understanding of the underlying mechanisms. This requires employing proper representations that capture key system properties while effectively simplify extraneous degrees of freedom.
Stochastic differential equations (SDE) provide such a flexible representation~\citep{arnold2014phenotypic,lande1976natural,chandrasekhar1943stochastic,nelson2004biological}, by representing the dominant system forces in the deterministic part of the equation, (\emph{drift function}) $f(\cdot):R^d \rightarrow R^d$, and summarising the unresolved or irrelevant degrees as stochastic forces acting on the system (\emph{diffusion}). The resulting evolution equation has the form
\begin{equation}\label{eq:system}
 \small \text{d}X_t = f(X_t) \text{d}t  + \sigma \,\text{d}W_t, \qquad X_0 = x_0 , \end{equation}
where $\sigma \in \mathcal{R}^{d\times d}$ stands for the noise amplitude, and $W_t$ for the d-dimensional vector of independent Wiener processes. The equation should be interpreted according to the Ito formalism. We observe the system state at distinct points in time through ${\mathcal{O}_k = \psi(X_{k \tau})}$, where ${X}_{k\tau}\dot{=}{X}_t\vert{_{t=\tau k}}$, with ${k =1, 2, \dots, K}$ observations collected at \textbf{inter-observation intervals} $\tau$, and want approximate the drift function $f(\cdot)$ from the observations $\{\mathcal{O}_k\}^K_{k=0}$. Here, we will consider $\psi(x)=x$, but the method generalises for monotonic functions $\psi(\cdot)$.


\section{Identifying stochastic systems from sparse-in-time observations}

For small inter-observation intervals $\tau$, we consider that we observe the continuous path of the system state $X_{0:T}$. \marginnote{\centering{\textbf{High-frequency observations}}}[-0cm] In that case, we can identify the drift function $f(\cdot)$ by the first order Kramers-Moyal coefficient~\citep{kramers1940brownian, moyal1949stochastic,tabar2019kramers} by empirically estimating conditional expectations of state increments~\citep{friedrich2000extracting,ragwitz2001indispensable, boninsegna2018sparse,siegert1998analysis}. For the Bayesian non-parametric counterpart of this approach, \citep{ruttor2013approximate, batz2018approximate} consider that transition probabilities between observations are Gaussian for $\text{d}t\rightarrow 0$, resulting in a (Gaussian) likelihood for the drift (see Sec.~\ref{appsec:b} Eq.~\ref{apeq:SDE_likelihood})
\begin{equation} \label{eq:SDE_likelihood}
\small    \mathcal{L}(X_{0:T}\mid f) = \exp \left[ -\frac{1}{2} \int^T_0 \| f(X_t)\|_{\sigma^2}^2  \text{d}t + \int^T_0 \langle f(X_t), X_{t+\text{d}t}-X_t \rangle \text{d}t \right].
\end{equation}
To identify the drift ~\cite{ruttor2013approximate} impose a Gaussian process prior on the function values $f$ (Eq.~\ref{eq7:full_gp_mean}). In Eq.~\ref{eq:SDE_likelihood} we introduced the weighted inner product ${\langle u, v\rangle \dot{=} u^{\top} \cdot \sigma^{-2} v}$ and weighted norm $\|u \|_{\sigma^2}  \dot{=} u^{\top} \cdot \sigma^{-2} u$.

 However, as the inter-observation interval $\tau$ increases, the transition probabilities between consecutive observations cannot be considered Gaussian, and thus the likelihood
(Eq.~\ref{eq:SDE_likelihood}) assumed between two successive observations is no longer valid if Eq.~\ref{eq:system} is non-linear. \marginnote{\centering{\textbf{Low-frequency observations}}}[-0cm] 
The likelihood for the drift $ \PP(\{\mathcal{O}_k\}_{k=1}^{K}|f)$ for such settings takes the form of a \emph{path integral}
\begin{equation}\label{eq7:path_integral_likelihood}
 \small   \PP(\{\mathcal{O}_k\}_{k=1}^{K}\mid f)= \int \PP (\{\mathcal{O}_k\}_{k=1}^{K}, X_{0:T}\mid f) \mathcal{D}(X_{0:T}) = \int \PP(\{\mathcal{O}_k\}_{k=1}^{K}\mid X_{0:T}) \PP(X_{0:T}|f) \mathcal{D}(X_{0:T}),
\end{equation}
where $\{\mathcal{O}_k\}_{k=1}^{K}$ denotes the set of $K$ discrete time observations, 
$\PP(X_{0:T}|f)$ the prior path probability assuming the dynamics of Eq.~\ref{eq:system}, $\mathcal{D}(X_{0:T}) $ identifies the formal volume element on the path space, while $\PP(\{\mathcal{O}_k\}_{k=1}^{K}| X_{0:T})$ stands for the likelihood of observations given the latent path $X_{0:T}$.

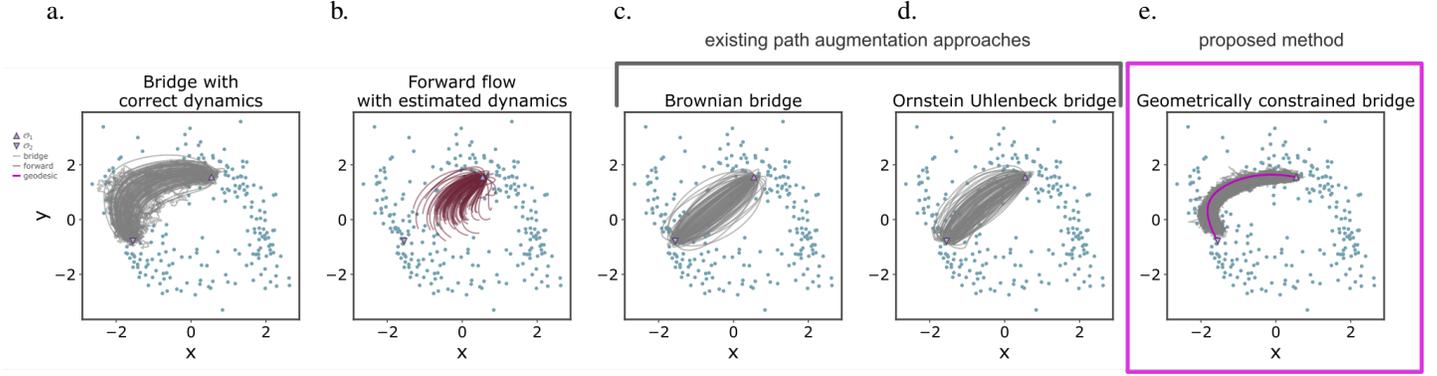
\begin{figure}
  \hspace{-80pt}
  \begin{overpic}[width=1.35\textwidth]{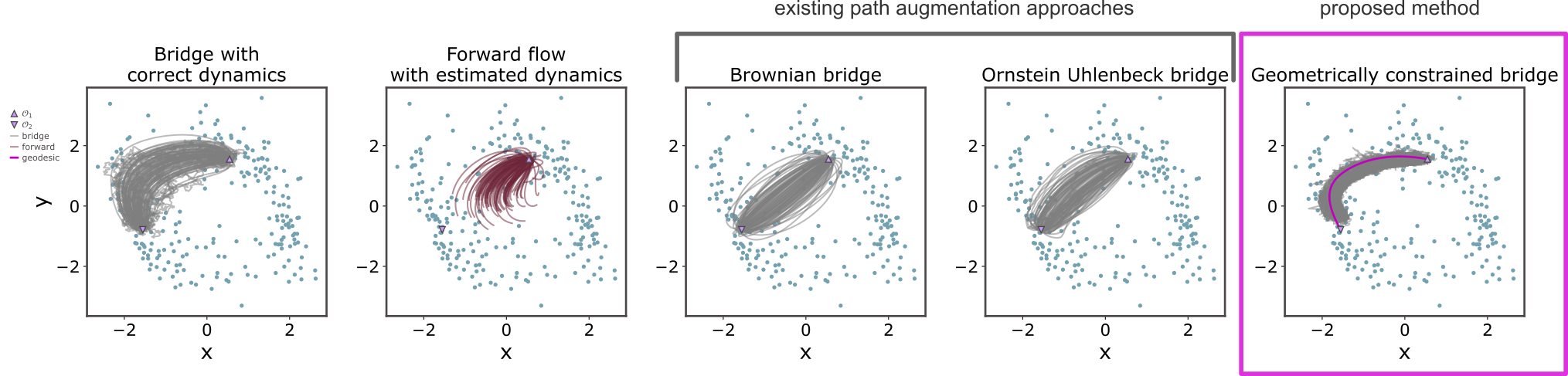}
  \put(3,25){a.}
  \put(23,25){b.}
  \put(43,25){c.}
  \put(63,25){d.}
  \put(80,25){e.}
  \end{overpic}
  \caption{ \textbf{Existing path augmentation strategies match poorly the underlying transition density between consecutive observations underestimating its curvature.} \textbf{a.)}  Stochastic bridge marginal density (\emph{grey}) between two successive observations $\mathcal{O}_1$ and $\mathcal{O}_2$ (\emph{pink triangles}) following the ground truth dynamics.
    \textbf{b.)}  Forward probability flow with estimated dynamics with a Gaussian likelihood (\emph{maroon}) matches poorly the correct transition density and often fails to reach the second observation $\mathcal{O}_2$ (\emph{downward pink triangle}).  Common path augmentation strategies employ either: \textbf{c.)} Brownian bridges, or \textbf{d.)} Ornstein Uhlenbeck (linear) bridge marginals resulting from local linearisations of the estimated drift with Gaussian likelihood. Both approaches match poorly the correct transition density, because they underestimate its curvature. \textbf{e.)} The proposed geometrically constrained path augmentation provides a better approximation of the underlying transition density by forcing the bridge paths towards the geodesic curve that connects consecutive observations on the manifold induced by the observations.     } 
  \label{fig:path_augmentations}
\end{figure}

From a geometric perspective, we consider that the nonlinear system dynamics induce an invariant density that may be approximated by a (possibly low dimensional) manifold. The sparse-in-time observations are samples of that manifold. For low-frequency observations, Euclidean distances employed for computing the state increments $X_{t+\tau}-X_t$ do not consider the geometry induced by the nonlinear dynamics, and thereby underestimate the curvature of the transition density between consecutive observations (Figure~\ref{fig:distances}).

\section{Geometric path augmentation}
Since the likelihood of Eq.~\ref{eq7:path_integral_likelihood} is intractable, we consider the unobserved continuous path between observations as latent random variables $X_{0:T}$, and obtain a maximum a posteriori estimate for the drift through Expectation Maximisation (EM)~\citep{dempster1977maximum}. Similar parametric~\citep{elerian2001likelihood,sermaidis2013markov} and non-parametric~\citep{batz2018approximate,ruttor2013approximate} methods have addressed the drift inference in the past, primarily in high-frequency observation settings. Our approach is based on the non-parametric method discussed in ~\citep{batz2018approximate,ruttor2013approximate}, with two significant advancements: \vspace{-6pt}
\begin{itemize}
    \item[\textbf{\textcolor{mymaroon2}{(i)}}]  We employ a path augmentation scheme following the \textbf{estimated nonlinear dynamics} resulting from inference with the Gaussian likelihood of Eq.~\ref{eq:SDE_likelihood} (as opposed to local linear approximations of these dynamics proposed in~\citep{batz2018approximate}).
    \item[\textbf{\textcolor{mymaroon2}{(ii)}}]  Importantly, we further \textbf{constrain the augmented paths to align with the} \textbf{geometry of the invariant density} between consecutive observations (Fig.~\ref{fig:distances} b.).
\end{itemize}

We follow an iterative algorithm, where at each iteration $n$ we perform the two following steps:\\
\textbf{(1.)}
     An \textbf{E(xpectation) step}, where given a drift estimate $\hat{f}^n$ we construct an approximate posterior over the latent variables $Q(X_{0:T}) \approx \PP(X_{0:T}|\{\mathcal{O}\}^K_{k=1}, \hat{f}^n(x))$.\\
\textbf{(2.)}  
     A \textbf{M(aximisation) step}, where we update the drift estimation.
     \vspace{-10pt}
\paragraph{$\bullet$ Approximate posterior over paths. (E-step)}
We approximate the continuous path trajectory $X_{0:T}$ between observations by a posterior path measure defined as the minimiser of the free energy \vspace{-6pt}
\begin{equation} \label{eq:free_energy2}
 \small   \mathcal{F}[Q] = \frac{1}{2} \int \limits^T_0 \int \Big[ \|g(x,t) - \hat{f}(x)\|_{\sigma^2}^2 + {U_{\mathcal{O}}(x,t)} +{U_{\mathcal{G}}(x,t)}   \Big] \, q_t(x)\, \text{d}x \,\text{d}t.
\end{equation}
The term ${U_{\mathcal{O}}(x,t) \dot{=} - \sum_{t_k} \ln \PP(\mathcal{O}_k|x) \delta(t- t_k)}$ forces the latent path to pass through the observations (or close to them depending on the observation process), while ${U_{\mathcal{G}}(x,t) \dot{=} \| \Gamma_t -x \|^2 }$ guides the latent path towards the geodesic curves $\gamma^k_{t'}$ that connect consecutive observations on the manifold $\mathcal{M}$ induced by the system's invariant density (Sec.~\ref{appsec:with}). Here we denote $\Gamma_t\dot{=} \{\gamma^k_{t'}\}_{t=(k-1)\tau+t' \tau}$, where $\gamma^k_{t'}$ is the geodesic connecting $\mathcal{O}_k$ and $\mathcal{O}_{k+1}$, and $t'\in [0,1]$. We identify the geodesic $\gamma^k_{t'}$ for each interval by learning the local metric of the manifold $\mathcal{M}$ (see Sec.~\ref{appsec:with} and~\cite{arvanitidis2019fast}). 

  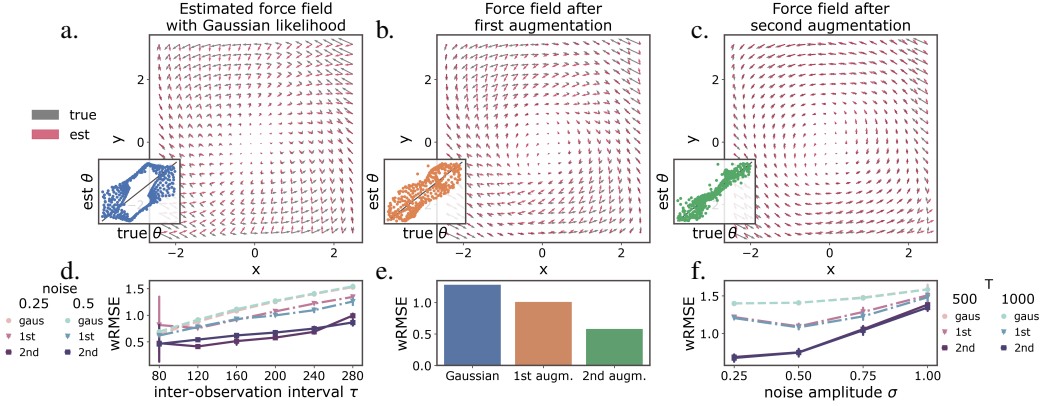
\begin{figure}
  \begin{overpic}[width=1.0\textwidth]{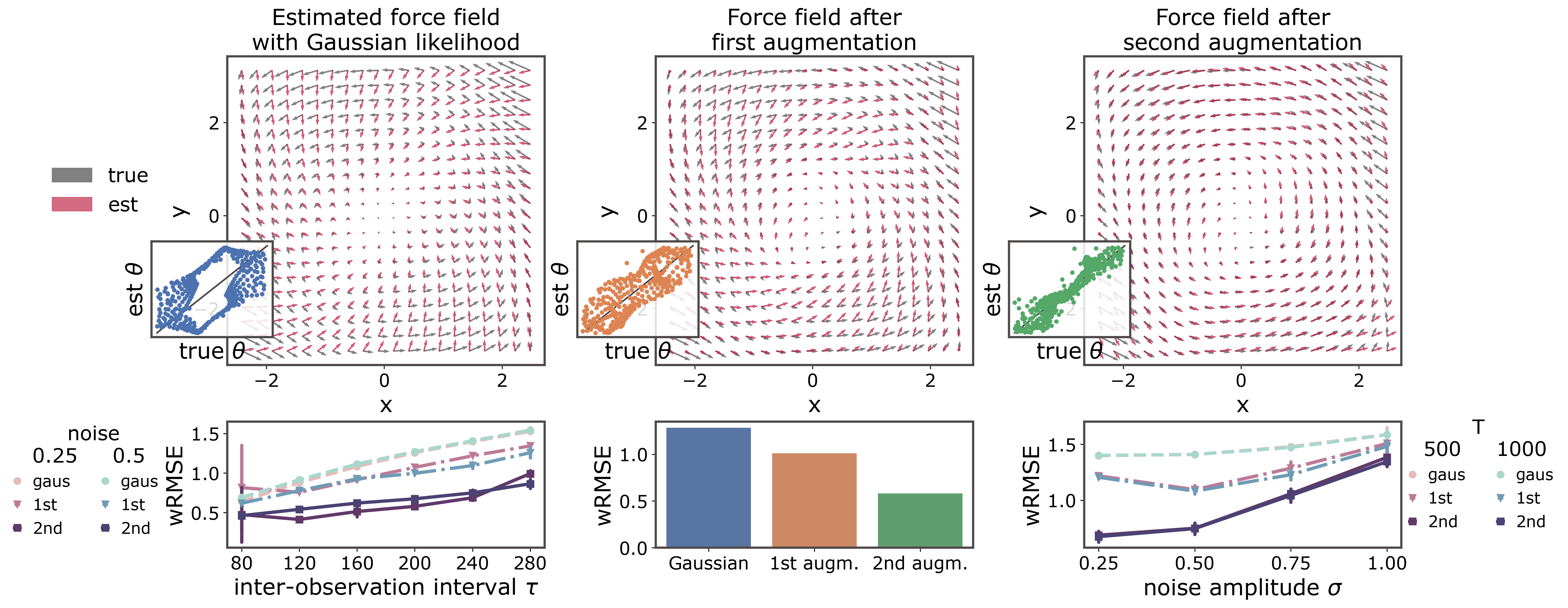}
  \put(6,35){a.}
  \put(36,35){b.}
  \put(66,35){c.}
  \put(6,12){d.}
  \put(36,12){e.}
  \put(66,12){f.}
  \end{overpic}
  \caption{ \textbf{Proposed path augmentation after two iterations already provides a good approximation of underlying drift.} Estimated (\emph{red}) and true (\emph{grey}) force field with \textbf{a.)} Gaussian likelihood   
    \textbf{b.)} after one and \textbf{c.)} after second iteration of augmentations. (\textbf{insets}) Ground truth against estimated angles for each point on the two dimensional grid. \textbf{e.)} Weighted root mean square error (wRMSE) for estimated drifts after each iteration for the presented example. The weights for averaging the error at each grid point are obtained from a kernel density estimation on the observations $\{\mathcal{O}_k\}^K_{k=1}$. \textbf{d.)} wRMSE against inter-observation interval $\tau$ for different noise conditions $\sigma=\{0.25,0.5\}$ for drift estimated with a Gaussian likelihood (\emph{gaus}-circles), after first augmentation (\emph{1st}-triangles), and after second augmentation (\emph{2nd}-squares) for $T=500$. \textbf{f.)} wRMSE against noise amplitude $\sigma$ in the system for different trajectory durations $T=\{500,1000\}$ time units for inter-observation interval $\tau=240$. Markers follow the same coding as in d.). Errorbars indicate one standard deviation over $5$ independent realisations.    } 
  \label{fig:res} \vspace{-6pt}
\end{figure}

Following~\cite{opper2019variational}, for each inter-observation interval $[\mathcal{O}_k, \mathcal{O}_{k+1}]$ we identify the posterior path measure (minimiser of Eq.~\ref{eq:free_energy2}) by the solution of a stochastic optimal control problem~\cite{maoutsa2022deterministic,maoutsa2021deterministica, maoutsa2022revealing} with the objective to obtain a time-dependent drift adjustment $u(x,t):=g(x,t) - \hat{f}(x)$ for the system with drift $\hat{f}(x)$ with {\textbf{initial and terminal constraints}} determined by $U_{\mathcal{O}}(x,t)$, and additional {\textbf{path constraints}} $U_{\mathcal{G}}(x,t)$.
\vspace{-10pt}
\paragraph{$\bullet$ Drift estimation. (M-step)} To estimate the drift from a sampled latent path, we assume a Gaussian process prior over function values and employ a sparse kernel approximation similar to~\cite{batz2018approximate} (see Sec.~\ref{appsec:drift_inference} for details).
\vspace{-10pt}

  \begin{SCfigure}
  \centering
  \begin{overpic}[width=0.65\textwidth]{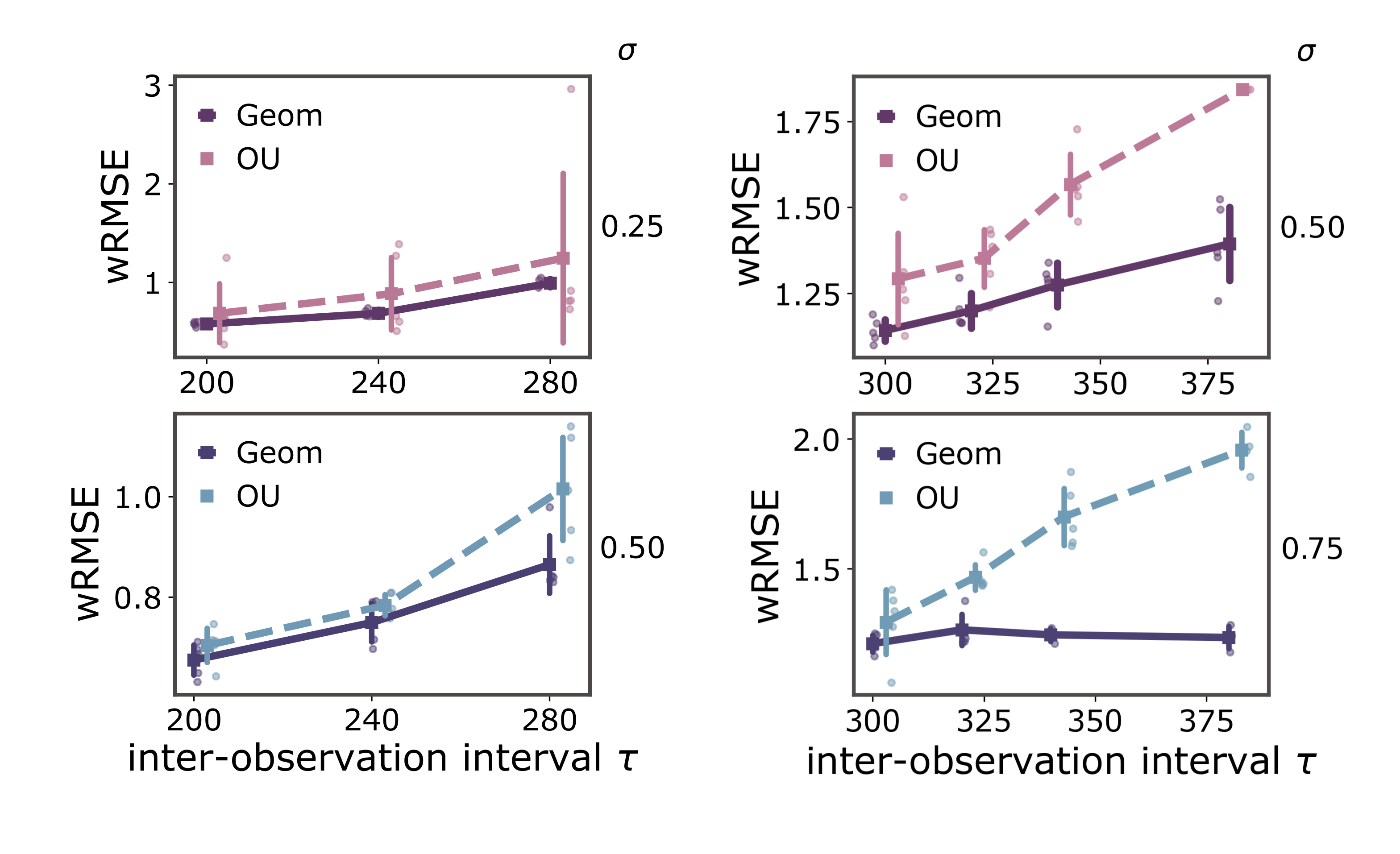}
  \put(6,55){a.}
  \put(6,30){b.}  
  \put(50,55){c.}
  \put(50,30){d.}
  \end{overpic}
  \caption{ \textbf{Comparison of proposed path augmentation with Ornstein-Uhlenbeck augmentation.}   Weighted root mean square error (wRMSE) against noise amplitude $\sigma$ for different inter-observation intervals for noise amplitude for moderate inter-observation intervals with \textbf{a.)} $\sigma=0.25$, and \textbf{b.)} $\sigma=0.50$, and for large inter-observation intervals with \textbf{c.)} $\sigma=0.50$, and \textbf{d.)} $\sigma=0.75$ and known direction of movement. In \textbf{c., d.} the inter-observation interval results in only one observation per oscillation period. 
  } 
  \label{fig:ou} 
\end{SCfigure}
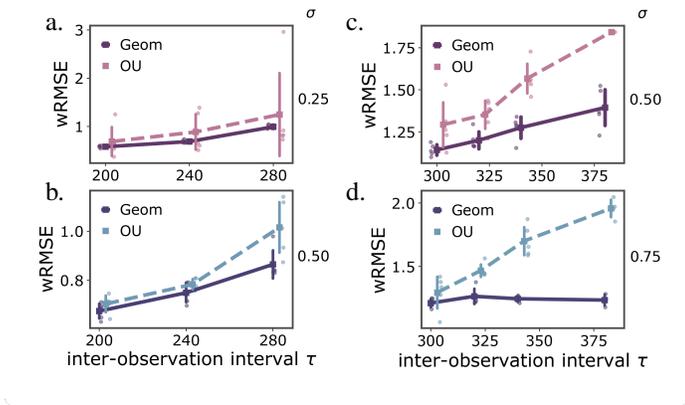

\section{Numerical experiments}\vspace{-5pt}
To demonstrate the performance of the proposed method we performed systematic estimations for a two-dimensional
Van der Pol oscillator under different noise conditions $\sigma$, observed at different inter-observation intervals $\tau$ for different lengths of trajectories $T$ (see Sec.~\ref{sec:details}). For the examined noise amplitudes (Fig.~\ref{fig:res} f.), the proposed path augmentation algorithm improves the naive estimation with Gaussian assumptions within two iterations for most noise amplitudes (Fig.~\ref{fig:res}). For increasing noise the improvement contributed by our approach decreases (Fig.~\ref{fig:res}f.), but is nevertheless not negligible.
 For low noise conditions, geodesics approximate the manifold structure better, however the path integral control is limited by the control costs proportional to inverse noise covariance. Our framework had comparable accuracy for all inter-observation lengths, but improvement was small for small lengths since in that setting the estimation with Gaussian likelihood already provides a good approximation of the ground truth drift. 
The proposed approach was compared to the augmentation method proposed by~\cite{batz2018approximate} (augmentation with Ornstein-Uhlenbeck bridges) and delivered more accurate estimates for larger inter-observation intervals. For inter-observation intervals with only one observation per oscillation period (Figure~\ref{fig:ou}c.,d.), the proposed approach delivered better results by considering additionally knowledge of the direction of movement in the state space (c.f. Sec.~\ref{sec:details}). The variance of estimates of the proposed method was smaller compared to Batz et al. due to conditioning on the invariant geometry of the system.

\vspace{-10pt}
\section{Conclusion and Discussion}\vspace{-5pt}

We introduced a new method for identifying stochastic systems from sparse-in-time observations of the system's state that reconciles approaches that rely purely on the temporal structure of the observations with those that approximate the geometry of the invariant density. Our method employs a path augmentation strategy that uses the nonlinear dynamics of a coarse drift estimate and further constrains the augmented paths to follow the local geometry of the system's invariant density. We found that the proposed approach provides efficient recovery of the underlying drift function for periodic or quasi-periodic systems under several noise conditions.
Only a limited number of inference methods have attempted to merge geometric and temporal perspectives for the identification of stochastic systems, such as the Langevin regression~\citep{callaham2021nonlinear}, TrajectoryNet~\citep{tong2020trajectorynet}, and the diffusion map method of~\cite{shnitzer2020manifold, shnitzer2016manifold}. However, our method differs from these approaches by employing the geodesic approximation of the underlying data geometry (see also Sec.~\ref{future} \textcolor{magenta}{Future directions}).

\newpage

\section*{Acknowledgements}
We thank Mina Miolane for early guidance on extracting geodesics from variational autoencoder based approximation of data manifolds, Stefan Sommer for answering questions on diffusions on manifolds, and Prof. Manfred Opper for prompting us to work on this problem. We further thank Georgios Arvanitidis for maintaining an open repository with his previous work on approximating geodesics from data manifolds. An earlier account of this work has been presented at the NeurIPS 2022 workshop \emph{Machine Learning and the Physical Sciences}~\citep{maoutsa2023geometric}. We further acknowledge that previous work from the Python~\citep{van1995python}, numpy~\citep{harris2020array}, scipy~\citep{2020SciPy-NMeth}, matplotlib~\citep{Hunter:2007}, seaborn~\citep{waskom2021seaborn}, GPflow~\citep{GPflow2017}, pyEMD~\citep{pele2009}, and pytorch~\citep{paszke2017automatic} communities facilitated the implementation of the computational part of this work.

An implementation of this work can be found in the following repository: \href{https://github.com/dimitra-maoutsa/Geometric-path-augmentation-for-SDEs}{https://github.com/dimitra-maoutsa/Geometric-path-augmentation-for-SDEs}.

\bibliography{iclr2023_workshop}
\bibliographystyle{iclr2023_workshop}


\newpage
\appendix
\appendixpage

\if False
{
\startcontents[sections]
\printcontents[sections]{l}{1}{\setcounter{tocdepth}{3}}
}

\fi

\section{Drift inference for high and low frequency observations}\label{appsec:b}
We consider systems whose evolution is captured by the stochastic differential equation Eq.~\ref{eq:system}.
\paragraph{High frequency observations.} When the system path $X_{0:T}$ is observed in continuous time, the infinitesimal transition probabilities of the diffusion process between consecutive observations are Gaussian, i.e.,
\begin{equation}
    \PP_f(X_{0:T} \mid f) \propto \exp \left( -\frac{1}{2 \text{d}t} \sum_t  \| X_{t+\text{d}t} -X_t - f(X_t)\text{d}t\|_{\sigma^2}^2     \right).
\end{equation}
In turn, the transition probability of (discretised) Wiener paths $P_{\mathcal{W}}(X_{0:T}) $ (i.e., paths from a drift-less process) can be expressed as
\begin{equation}
    \PP_{\mathcal{W}}(X_{0:T})=\exp \left( -\frac{1}{2 \text{d}t} \sum_t  \| X_{t+\text{d}t} -X_t\|_{\sigma^2}^2     \right),
\end{equation}
where $\|u \|_{\sigma^2}  \dot{=} u^{\top} \cdot {\sigma}^{-2} u$ denotes the weighted norm with $D\dot{=}\sigma^2$ indicating the noise covariance. 
We can thus express the likelihood for the drift $f$ by the Radon-Nykodym derivative between $P_f(X_{0:T}|f)$ and $\PP_{\mathcal{W}}(X_{0:T})$
 for paths $X_{0:T}$ within the time interval $[0,\,T]$
~\citep{liptser2013statistics}
\begin{equation} \label{apeq:SDE_likelihood}
    \mathcal{L}(X_{0:T} \mid f) = \exp \left[ -\frac{1}{2} \sum_t \| f(X_t)\|_{\sigma^2}^2  \text{d}t + \sum_t \langle f(X_t), X_{t+\text{d}t}-X_t \rangle_{\sigma^2}  \right],\end{equation}
where for brevity we have introduced the notation $\langle u, v\rangle \dot{=} u^{\top} \cdot {\sigma}^{-2} v$ for the weighted inner product with respect to the inverse noise covariance ${\sigma}^{-2}$. This expression results from applying the Girsanov theorem on the path measures induced by a process with drift $f$ and a Wiener process, with same diffusion $\sigma$, and employing an Euler-Maruyama discretisation on the continuous path $X_{0:T}$.

The likelihood given a continuously observed path of the SDE (Eq.~\ref{apeq:SDE_likelihood}) has a quadratic form in terms of the drift function. Therefore a Gaussian measure over function values (Gaussian process) is a natural conjugate prior for this likelihood. To identify the drift in a non-parametric form, we assume a Gaussian process prior for the function values $f \sim \PP_0({f}) =\text{GP}(m^f, k^f)$, where $m^f$ and $k^f$ denote the mean and covariance function of the Gaussian process ~\citep{ruttor2013approximate}. The prior measure can be written as
\begin{equation}
    \PP_0({f})  = \exp\left[-\frac{1}{2} \int \int f(x) \left(k^f(x,x') \right)^{-1}f(x') \text{d}x \text{d}x'\right],
\end{equation}
if we consider a zero mean Gaussian process $m^f=0$.

Bayesian inference for the drift function $f$ requires the computation of a probability distribution in the function space, the posterior probability distribution $\PP_f(f \mid X_{0:T})$. From the Bayes' rule the posterior can be expressed as 
\begin{equation}
    \PP_f(f\mid X_{0:T}) =\frac{ \PP_0(f) \mathcal{L}(X_{0:T} \mid f)}{Z} \propto \PP_0(f) \mathcal{L}(X_{0:T} \mid f),
\end{equation}
where $Z$ denotes a normalising factor defined as a path integral 
\begin{equation}
    Z = \int \PP_0(f) \mathcal{L}(X_{0:T} \mid f) \mathcal{D}f ,
\end{equation}
where $ \mathcal{D}f$ denotes integration over the Hilbert space $f: H_0[f] < \infty$ . Here we have expressed the prior probability over functions as $\PP_0(f) = e^{-H_0[f]}$.
In~\citep{ruttor2013approximate} the authors show that in the continuous time limit, nonparametric estimation of drift functions becomes equivalent to Gaussian process regression,  with the objective to identify the mapping from the system state $X_t$ to state increments $\text{d}X_t$~\citep{rasmussen2003gaussian}. More precisely, we consider $N$ observations of the system state $X_t$ as the regressor, with associated response variables 
\begin{equation}\label{eq:increments}
    Y_t = \frac{X_{t+\text{d}t}-X_t}{\text{d}t},
\end{equation} and denote the kernel function of the Gaussian process by $k(x,x')$. 

If we denote with $\mathcal{X}= \{X_t\}^{T-\text{d}t}_{t=0}$ and $\mathcal{Y}= \{Y_t\}^{T-\text{d}t}_{t=0}$ the set of state observations and observation increments, the mean of the posterior process over drift functions $f$ can be expressed as
\begin{equation}\label{eq7:full_gp_mean}
    \bar{f}(x) = k^f(x,\mathcal{X})^{\top} \left( \mathcal{K} + \frac{{\sigma}^{2}}{\text{d}t} I_N\right)^{-1} \mathcal{Y},
\end{equation}
where we abused the notation and denoted with $k^f(x, \mathcal{X})$ the vector resulting from evaluating the kernel $k^f$ at points $x$ and $\{\mathcal{O}_t\}^{K-1}_{k=1}$. Similarly $\mathcal{K} = k^f(\mathcal{X}, \mathcal{X})$ stands for the $(K-1)\times (K-1)$ matrix resulting from evaluation of the kernel on all observation pairs. 
In a similar vein, the posterior variance can be written as
\begin{equation} \label{eq7:full_gp_cov}
    \Sigma^2(x) = k^f(x,x) - k^f(x,\mathcal{X})^{\top} \left(\mathcal{K} + \frac{{\sigma}^{2}}{\text{d}t} \right)^{-1} k^f(x,\mathcal{X}),
\end{equation}
where the term ${\sigma}^{2}/\text{d}t$ plays the role of observation noise.

\paragraph{Low frequency observations.} When the inter-observation interval becomes large (\emph{low frequency observations}), the Gaussian likelihood of Eq.~\ref{apeq:SDE_likelihood} becomes invalid, since for large inter-observation intervals the transition density is no longer Gaussian. Thus, drift estimation with Gaussian assumptions ~\citep{friedrich1997description,ruttor2013approximate} becomes inaccurate. To mitigate this issue Lade~\citep{lade2009finite} introduced a method to compute finite time corrections for the drift estimates, which has been applied (to the best of our knowledge) mostly to one dimensional problems~\citep{honisch2011estimation}. On the other hand, the statistics community has proposed path augmentation schemes that augment the observed trajectory to a nearly continuous-time trajectory by sampling a simplified system's dynamics between observations~\citep{golightly2008bayesian,papaspiliopoulos2012nonparametric,sermaidis2013markov,beskos2006retrospective,chib2006likelihood}.
However for large inter-observation intervals and for nonlinear systems the simplified dynamics employed for path augmentation match poorly the underlying path statistics, and these methods show poor convergence rates or fail to identify the correct dynamics (Figure~\ref{fig:path_augmentations} c. and d.). We point out here, that path augmentation with Ornstein Uhlenbeck bridges using as drift the local linearisation of the \textbf{correct} dynamics, provides a good approximation of the underlying transition density. However, during inference, the true underlying dynamics are unknown, and the proposed local linearisations on inaccurate drift estimates~\citep{batz2018approximate} perform poorly for low frequency observations.

Notice that as the inter-observation interval $\tau$ increases, the Gaussian likelihood assumed between two successive observations is no longer valid if the system is non-linear or when the noise is state dependent. 
The likelihood for the drift for such settings can be expressed in terms of a \emph{path integral}
\begin{equation}\label{apeq7:path_integral_likelihood}
    \PP(\mathcal{O}_{1:K}\mid f) = \int \PP(\mathcal{O}_{1:K}\mid X_{0:T}) \PP(X_{0:T}\mid f) \mathcal{D}(X_{0:T}),
\end{equation}
where $\mathcal{O}_{1:K}\dot{=}\{\mathcal{O}_k\}_{k=1}^{K}$ denotes the set of $K$ discrete time observations, 
$\PP(X_{0:T}\mid f)$ the prior path probability resulting from a diffusion process with drift $f(x)$, $\mathcal{D}(X_{0:T}) $ identifies the formal volume element on the path space, and $\PP(\mathcal{O}_{1:K}\mid X_{0:T})$ stands for the likelihood of observations given the latent path $X_{0:T}$.

However, the path integral of Eq.~\ref{apeq7:path_integral_likelihood} is intractable for nonlinear systems, thus we need to simultaneously estimate the drift and latent state of the diffusion process, i.e., to approximate the joint posterior measure of latent paths and drift functions $\PP(X_{0:T},f \mid \mathcal{O}_{1:K})$. Therefore we consider the unobserved continuous path $X_{0:T}$ as latent random variables and employ an Expectation Maximisation (EM) algorithm to identify a maximum a posteriori estimate for the drift function. 
More precisely, we follow an iterative algorithm, where at each iteration $n$ we alternate between the two following steps:

\begin{multicols}{2}
\begin{minipage}[c]{0.95\linewidth}
     An \textbf{Expectation} step, where given a drift estimate $\hat{f}^n(x)$ we construct an approximate posterior over the latent variables ${Q(X_{0:T}) \approx \PP(X_{0:T}\mid \mathcal{O}_{1:K}, \hat{f}^n(x))}$, and compute the expected log-likelihood of the augmented path 
    \begin{equation} \label{eq:estep}
    \mathfrak{L}\big(\hat{f}^n(x), Q\big) = \mathbb{E}_Q\Big[\ln \mathscr{L}\big(X_{0:T}\mid \hat{f}^n(x)\big) \Big].
    \end{equation}
    \end{minipage}
    
     A \textbf{Maximisation} step, where we update the drift estimation by maximising the expected log likelihood
    \begin{equation} \label{eq:mstep}
        f^{n+1}(x) = \arg \max_f \Big[\mathfrak{L}\big(f^n(x),Q\big)-\ln \PP_0\big(f^n(x)\big) \Big].
    \end{equation}
\end{multicols}
In Eq.~\ref{eq:mstep} $\PP_0$ denotes the Gaussian process prior over function values.

\subsection{Approximate posterior over paths.} 

Here we first formulate the approximate posterior over paths (conditional distribution for the path given the observations) by considering only individual observations as constraints (Section~\ref{appsec:without}). However, this approach results computationally taxing calculations during path augmentation, since the observations are atypical states of the initially estimated drift. To overcome this issue, we subsequently extend the formalism (Section~\ref{appsec:with}) to incorporate constraints that consider also the local geometry of the observations.

\subsubsection{Approximate posterior over paths \underline{without} geometric constraints.}\label{appsec:without}
Given a drift function (or a drift estimate) $\hat{f}(x)$ we can apply variational techniques to approximate the posterior measure over the latent path conditioned on the observations $\{\mathcal{O}_k\}^K_{k=1}$. We consider that the prior process (the process without considering the observations $\{\mathcal{O}_k\}^K_{k=1}$) is described by the equation
\begin{equation} \label{appeq:prior_sde}
  \PP(X_{0:T}\mid \hat{f}): \qquad  \text{d}X_t = \hat{f}(X_t) \text{d}t + \sigma \text{d}W_t.
\end{equation}
We will define an approximating (posterior) process that is conditioned on the observations. The conditioned process is also a diffusion process with the same diffusion as Eq.~\ref{appeq:prior_sde} but with a modified, time-dependent drift $g(x,t)$ that accounts for the observations~\citep{chetrite2015variational,majumdar2015effective}. 
We identify the approximate posterior measure $Q$ with the posterior measure induced by an approximating process that is conditioned by the observations $\mathcal{O}_{1:K}$~\citep{opper2019variational},
with governing equation
\begin{equation} \label{appeq:sde_q}
    Q(X_{0:T}): \qquad \text{d}X_t = g(X_t,t) \text{d}t + \sigma \text{d}W_t=\left(\hat{f}(X_t) + \sigma^2 u(X_t,t) \right) \text{d}t +  \sigma \text{d}W_t.
\end{equation}

The effective drift $g(X_t,t)$ of Eq.~\ref{appeq:sde_q} may be obtained from the solution of the variational problem of minimising the free energy
\begin{equation}\label{eq:free_energy}
    \mathcal{F}[Q] = \mathcal{KL}\Big(Q(X_{0:T})||\PP(X_{0:T}\mid \hat{f}) \Big)- \sum \limits_{k=1}^K {E}_Q[ \ln \PP(\mathcal{O}_{k}\mid X_{t_k})].
\end{equation}

By applying the Cameron-Girsanov-Martin theorem we can express the Kullback-Leibler divergence between the two path measures induced by the diffusions with drift $\hat{f}(x)$ and $g(x,t)$ as
\begin{align}
    \mathcal{KL}\Big(Q(X_{0:T})||\PP(X_{0:T}|\hat{f}) \Big) &= E_{{Q}}\left[\text{ln}\left(\frac{d {Q}(X_{0:T})}{d {P} \left( X_{0:T} \vert \hat{f}  \right)} \right)\right] \\ 
    &=E_{{Q}}\left[  \left( - \frac{1}{2}  \int_0^T { { \|\hat{f}(X_t)-g(X_t,t) \|_{\sigma^{2}}^2}\text{d}t} + \int_0^T {\frac{ \hat{f}(X_t)-g(X_t,t)  }{{\sigma^{2}}} \text{d}W_t} \right)\right]\\
    &=E_{{Q}}\left[  \left( - \frac{1}{2}  \int_0^T { { \|\hat{f}(X_t)-g(X_t,t) \|_{\sigma^{2}}^2}\text{d}t} +V_T \right)\right]\\
    &= \frac{1}{2} \int \limits^T_0 \int \| g(x,t) -\hat{f}(x)   \|_{\sigma^{2}}^2 \, q_t(x)\, \text{d}x \, \text{d}t + \mathfrak{C} \label{eq:KL_1},
\end{align}
where $q_t(x)$ stands for the marginal density for $X_t$ of the approximate process. In the third line we have introduced the random variable $V_T = \int_0^T {\frac{ \hat{f}(X_t)-g(X_t,t)  }{{\sigma^{2}}} \text{d}W_t}$. Under the assumption that the function ${\ell(X_t) = \hat{f}(X_t)-g(X_t,t)}$ is bounded, piece-wise continuous, and in $L^2[0,\infty)$ , $V_T$ follows the distribution $\mathcal{N}\left(V_T \mid 0, \int_0^T \ell^2(s) \text{d}s\right)$, which for a given $T$ will result into a constant $\mathfrak{C}$. Thus the second term in Eq.~\ref{eq:KL_1} is not relevant for the minimisation of the free energy and will be omitted. 

We can thus express the free energy of Eq.~\ref{eq:free_energy} as~\citep{opper2019variational}
\begin{equation} \label{appeq:free_energy2}
    \mathcal{F}[Q] = \frac{1}{2} \int \limits^T_0 \int \Big[ \|g(x,t) - \hat{f}(x)\|_{\sigma^{2}}^2 + U(x,t)   \Big] \, q_t(x)\, \text{d}x \,\text{d}t,
\end{equation}
where the term $U(x,t)$ accounts for the observations $U(x,t) = - \sum \limits_{t_k} \ln \PP(\mathcal{O}_k \mid x) \delta(t- t_k)$.

The minimisation of the functional of the free energy can be construed as a stochastic control problem~\citep{opper2019variational} with the objective to identify a time-dependent drift adjustment $u(x,t):=g(x,t) - \hat{f}(x)$ for the system with drift $\hat{f}(x)$ so that the controlled dynamics fulfil the constraints imposed by the observations.

For the case of exact observations, i.e., for an observation process $\psi(x) = x$, we can compute the drift adjustment for each of the $K-1$ inter-observation intervals independently. Thus for each interval between consecutive observations, we identify the optimal control $u(x,t)$ required to construct a stochastic bridge following the dynamics of Eq.~\ref{appeq:prior_sde} with initial and terminal states the respective observations $\mathcal{O}_k$ and $\mathcal{O}_{k+1}$. 

The optimal drift adjustment for such a stochastic control problem for the inter-observation interval between $\mathcal{O}_k$ and $\mathcal{O}_{k+1}$ can be obtained from the solution of the backward equation (see~\citep{maoutsa2022deterministic, maoutsa2021deterministica})
\begin{equation}
    \frac{\partial \phi_t(x)}{\partial t} = - \mathcal{L}_{\hat{f}}^{\dagger} \phi_t(x) + U(x,t) \phi_t(x),
\end{equation}
with terminal condition $\phi_T(x) = \chi(x) = \delta(x-\mathcal{O}_{k+1}) $ and with $\mathcal{L}_{\hat{f}}^{\dagger}$ denoting the adjoint Fokker-Planck operator for the process of Eq.~\ref{appeq:prior_sde}.
As shown in Maoutsa et al.~\citep{maoutsa2022deterministic, maoutsa2021deterministica} the optimal drift adjustment $u(x,t)$ can be expressed in terms of the difference of the logarithmic gradients of two probability flows
\begin{equation}
    u^*(x,t) = D \Big( \nabla \ln q_{T-t}(x) - \nabla \ln \rho_t(x) \Big),
\end{equation}
where $\rho_t$ fulfils the forward (filtering) partial differential equation (PDE)
 \begin{equation}
\frac{\partial \rho_t(x)}{\partial t} = {\cal{L}}_{\hat{f}} \rho_t(x) - U(x,t) \rho_t(x),
\label{eq:FPE2} 
\end{equation}
while $q_t$ is the solution of a time-reversed PDE that depends on the logarithmic gradient of $\rho_t(x)$
\begin{align}\label{Fokker_bridge3}
\frac{\partial {q}_{t}(x)}{\partial t} &= 
-\nabla\cdot \Bigg[\Big(\sigma^2\nabla \ln  \rho_{T-t} (x)  - f(x, T-t)\Big)  {q}_{t} (x)\Bigg] +  \frac{\sigma^2}{2} \nabla^2 {q}_{t} (x) , 
\end{align}
with initial condition ${q}_{0} (x) \propto \rho_T(x) \chi(x)$
.

\subsubsection{Approximate posterior over paths \underline{with} geometric constraints.}\label{appsec:with}

The previously described construction of the approximate measure in terms of stochastic bridges is relevant when the observations have non vanishing probability under the law of the prior diffusion process of Eq.~\ref{appeq:prior_sde}. However, when the prior process (with the estimated drift $\hat{f}$) differs considerably from the process that generated the observations, such a construction might either provide a bad approximation of the underlying path measure, or show slow numerical convergence in the construction of the diffusion bridges.
To overcome this issue, we consider here additional constraints for the posterior process that force the paths of the posterior measure to respect the local geometry of the observations. In the following we provide a brief introduction on the basics of Riemannian geometry and consequently continue with the geometric considerations of the proposed method.

\paragraph{Riemannian geometry.}
A $d$-dimensional \textbf{Riemannian manifold}~\citep{do1992riemannian,lee2018introduction} $\left(\mathcal{M}, h \right)$ embedded in a $D$-dimensional ambient space $\mathcal{X} = \mathcal{R}^D$ is a smooth curved $d$-dimensional surface 
endowed with a smoothly varying inner product (Riemannian) \textbf{metric} $h: x \rightarrow \langle \cdot | \cdot \rangle_x$ on $\mathcal{T}_x\mathcal{M}$. A tangent space $\mathcal{T}_x \mathcal{M}$ is defined at each point $x \in \mathcal{M}$. The Riemannian metric $h$ defines a canonical volume measure on the manifold $\mathcal{M}$. Intuitively this characterises how to compute inner products locally between points on the tangent space of the manifold $\mathcal{M}$, and therefore determines also how to compute norms and thus distances between points on $\mathcal{M}$.

A \textbf{coordinate chart} $(G,\phi)$ provides the mapping from an open set $G$ on $\mathcal{M}$ to an open set $V$ in the Euclidean space. The dimensionality of the manifold is $d$ if for each point $x\in \mathcal{M}$ there exists a local neighborhood $G \subset  \mathcal{R}^d$.
We can represent the metric $h$ on the local chart $(G,\phi)$ by the positive definite matrix (\textbf{metric tensor}) $H(x) = (h_{i,j})_{x, 0 \leq i,j,\leq d} = \left( \langle  \frac{\partial}{\partial x_i}|  \frac{\partial}{\partial x_j}\rangle_x  \right)_{0 \leq i,j,\leq d}$ at each point $x \in G$.

For $v,w \in \mathcal{T}_x\mathcal{M}$ and $x \in G$, their inner product can be expressed in terms of the matrix representation of the metric  $h$ on the tangent space $\mathcal{T}_x\mathcal{M}$ as $\langle v|w \rangle_x = v^{\top} H(x)w$, where $H(x)\in \mathcal{R}^{d \times d}$ .

The \textbf{length of a curve} $\gamma:[0,1]\rightarrow \mathcal{M}$ on the manifold is defined as the integral of the norm of the tangent vector 
\begin{equation}\label{eq:ell}
\ell(\gamma_{t'}) = \int^1_0\| \dot{\gamma}_{t'}\|_h \text{d}t' = \int^1_0 \sqrt{ \dot{\gamma}_{t'}^{\top} H(\gamma_{t'}) \dot{\gamma}_{t'}    } \text{d}t',\end{equation}
where the dotted letter indicates the velocity of the curve $\dot{\gamma}_{t'}=\partial_{t'} \gamma_{t'}$. A \textbf{geodesic curve} is a locally length minimising smooth curve that connects two given points on the manifold.

\paragraph{Riemannian geometry of the observations.} 
For approximating the posterior over paths we take into account the geometry of the invariant density as it is represented by  the observations.  
To that end, we consider systems whose dynamics induce invariant (inertial) manifolds that contain the global attractor of the system and on which system trajectories concentrate~\citep{wiggins1994normally,mohammed1999stable, girya1995inertial, fenichel1971persistence, arnold1990stochastic, carverhill1985flows}. We assume thus that the continuous-time trajectories $X_{0:T} \in \mathcal{R}^d$ of the underlying system concentrates on an invariant manifold $\mathcal{M} \in \mathcal{R}^{m \leq d}$ of dimensionality $m$ (possibly) smaller than $d$.
The discrete-time observations $\mathcal{O}_k$ are thus samples of the manifold $\mathcal{M}$.
The central premise of our approach is that \textbf{unobserved paths between successive observations will be lying either \emph{on} or \emph{in the vicinity} of the manifold} $\mathcal{M}$. In particular, we postulate that unobserved paths should lie \textbf{in the vicinity of geodesics that connect consecutive observations} on $\mathcal{M}$. To that end we propose a path augmentation framework that constraints the augmented paths to lie in the vicinity of identified geodesics between consecutive observations.

However, while this view of a lower dimensional manifold embedded in a higher dimensional ambient space helps to build our intuition for the proposed method, for computational purposes we adopt a complementary view inspired by the discussion in~\citep{frohlich2021bayesian}. According to this view, we consider the entire observation space $\mathcal{R}^d$ as a smooth Riemannian manifold, $\mathcal{M}\dot{=}\mathcal{R}^d$, characterised by a Riemannian metric $h$. The effect of the nonlinear geometry of the observations is then captured by the metric $h$. Thus to approximate the geometric structure of the system's invariant density, we learn the Riemannian metric tensor $H:\mathcal{R}^d \rightarrow \mathcal{R}^{d \times d} $ and compute the geodesics between consecutive observations according to the learned metric. Intuitively according to this view the observations $\{\mathcal{O}_k\}^K_{k=1}$ introduce distortions in the way we compute distances on the state space.

In effect this approach does not reduce the dimensionality of the space we operate, but changes the way we compute inner products and thus distances, lengths, and geodesic curves on $\mathcal{M}$. The alternative perspective of working on a lower dimensional manifold would strongly depend on the correct assessment of the dimensionality of said manifold. For example, one could use a Variational Autoencoder to approximate the observation manifold and subsequently obtain the Riemannian metric from the embedding of the manifold mediated by the decoder. 
However, our preliminary results of such an approach revealed that such a method requires considerable fine tuning to adapt to the characteristics of each dynamical system and is sensitive to the estimation of the dimensionality of the approximated manifold. 

To learn the Riemannian metric and compute the geodesics we follow the framework proposed by~\cite{arvanitidis2019fast}.
In particular, we approximate the local metric induced by the observations at location $\mathbf{x}$ of the state space, in a non-parametric form by the inverse of the weighted local diagonal covariance computed on the observations as~\citep{arvanitidis2019fast}
\begin{equation}
    H_{dd}(\mathbf{x}) = \left(  \sum\limits^K_{i=1} w_i(\mathbf{x}) \left( x^{(d)}_i - x^{(d)}\right)^2 + \epsilon   \right)^{-1},
\end{equation}
with weights $w_i(\mathbf{x}) = \exp \left(- \frac{\|  \mathbf{x}_i - \mathbf{x} \|^2_2}{2 \sigma^2_{\mathcal{M}}}  \right)$, and $x^{(d)}$ denoting the $d$-th dimensional component of the vector $\mathbf{x}$. The parameter $\epsilon > 0$ ensures non-zero diagonals of the weighted covariance matrix, while $\sigma_{\mathcal{M}}$ characterises the curvature of the manifold.

Between consecutive observations for each interval $[\mathcal{O}_k, \mathcal{O}_{k+1}]$, we identify the geodesic $\gamma^k_{t'}$ as the energy minimising curve, i.e., as the minimiser of the kinetic energy functional $\mathcal{E}(\gamma^k_{t'}) =\int^1_0 L_{{\mathcal{M}}}(\gamma^k_{t'}, \dot{\gamma}^k_{t'}) \text{d}t'$
\begin{equation} \nonumber
  \gamma^{k*}_{t'} =  \underset{\gamma^k_{t'}, \gamma^k_0 = \mathcal{O}_k, \gamma^k_1=\mathcal{O}_{k+1}}{\arg\min} \int^1_0 L_{{\mathcal{M}}}(\gamma^k_{t'}, \dot{\gamma}^k_{t'}) \text{d}t',
\end{equation}

\begin{equation} \label{eq:machlup}
\text{with} \;\;\;\;  \int^1_0 L_{{\mathcal{M}}}(\gamma^k_{t'}, \dot{\gamma}^k_{t'}) \text{d}t'= \frac{1}{2}  \int^1_0 \|\dot{\gamma}^k_{t'} \|^2_h  ,
\end{equation} \label{eq:geodesic_lagrangian}
where $L_{{\mathcal{M}}}(\gamma^k_{t'}, \dot{\gamma}^k_{t'})$ denotes the Lagrangian.
The minimising curve of this functional is the same as the minimiser of the curve length functional $\ell(\gamma_{t'})$ (Eq.~\ref{eq:ell}), i.e., the geodesic~\citep{do1992riemannian}.

By applying calculus of variations, the minimising curve of the functional $\mathcal{E}(\gamma^k_{t'}) $ can be obtained from the Euler-Lagrange equations, resulting in the following system of second order differential equations~\citep{arvanitidis2017latent,do1992riemannian}
\begin{equation}\label{eq:geode}
    \ddot{\gamma_t}^k = -\frac{1}{2} {H(\gamma^k_t)}^{-1} \Bigg( 2 \left( I \otimes (\dot{\gamma_t}^k)^{\top} \right) \frac{\partial \text{vec}[H(\gamma^k_t)]}{\partial \gamma^k_t}  \dot{\gamma_t}^k  - \frac{\partial \text{vec}[H(\gamma^k_t)]^{\top}}{\partial \gamma^k_t} \left(\dot{\gamma_t}^k \otimes \dot{\gamma_t}^k  \right)\Bigg),
\end{equation}
with boundary conditions $\gamma^k_0 = \mathcal{O}_k $ and $ \gamma^k_1=\mathcal{O}_{k+1}$,
where $\otimes$ stands for the Kroenecker product, and $\text{vec}[A]$ denotes the vectorisation operation of matrix $A$ through stacking the columns of $A$ into a vector.
\cite{arvanitidis2019fast} obtain the geodesics by approximating the solution of the boundary value problem of Eq.~\ref{eq:geode} with a probabilistic differential equation solver.

\paragraph{Extended free energy functional.} We denote the collection of individual geodesics by $\Gamma_t\dot{=} \{\gamma^k_{t'}\}_{t=(k-1)\tau+t' \tau}$, where $\gamma^k_{t'}$ is the geodesic connecting $\mathcal{O}_k$ and $\mathcal{O}_{k+1}$, and $t'\in [0,1]$ denotes a rescaled time variable. Additional to the constraints imposed in the previously explained setting (Sec~\ref{appsec:without}), here we add an extra term in the free energy ${U_{\mathcal{G}}(x,t) \dot{=} \| \Gamma_t -x \|^2 }$ that accounts for the local geometry of the invariant density, and guides the latent path towards the geodesic curves $\gamma^k_{t'}$ that connect consecutive observations 
\begin{equation} \label{apeq:free_energy2}
 \small   \mathcal{F}[Q] = \frac{1}{2} \int \limits^T_0 \int \Big[ \|g(x,t) - \hat{f}(x)\|_{\sigma^2}^2 + U_{\mathcal{O}}(x,t) + \beta U_{\mathcal{G}}(x,t)   \Big] \, q_t(x)\, dx \,\text{d}t.
\end{equation}
Here we denote the observation term by $U_{\mathcal{O}}(x,t) \dot{=} - \sum_{t_k} \ln \PP(\mathcal{O}_k|x) \delta(t- t_k)$, while $\beta$ stands for a weighting constant that determines the relative weight of the geometric term in the control objective.

 Following~\cite{opper2019variational}, for each inter-observation interval $[\mathcal{O}_k, \mathcal{O}_{k+1}]$ we identify the posterior path measure (minimiser of Eq.~\ref{apeq:free_energy2}) by the solution of a stochastic optimal control problem~\citep{maoutsa2022deterministic, maoutsa2021deterministica} with the objective to obtain a time-dependent drift adjustment $u(x,t):=g(x,t) - \hat{f}(x)$ for the system with drift $\hat{f}(x)$ with initial and terminal constraints defined by $U_{\mathcal{O}}(x,t)$, and additional path constraints $U_{\mathcal{G}}(x,t)$.

\subsection{Approximate posterior over drift functions.}\label{appsec:drift_inference}

For a fixed path measure ${Q}$, the optimal measure for the drift ${Q}_f$ is a Gaussian process given by
\begin{equation} \label{appeq:drift_measure}
    {Q}_f \propto \PP_f \exp\left({ -\frac{1}{2} \int  \|f(x)\|_{\sigma^{2}}^2 A(x) - 2 \langle f(x), B(x) \rangle_{\sigma^{2}}  } \text{d}x\right),
\end{equation}
with $$A(x)\dot{=} \int^T_{0} p_t(x) \text{d}t,$$ and $$B(x)\dot{=} \int^T_{0} p_t(x) g(x,t) \text{d}t, $$ where $p_t(x)$ denotes the marginal constrained density of the state $X_t$. The function $g(x,t)$ denotes the effective drift.

 We assume a Gaussian process prior for the unknown function $f$, i.e., $f \sim \PP_0({f}) =\text{GP}(m^f, k^f)$ where $m^f$ and $k^f$ denote the mean and covariance function of the Gaussian process. Following Ruttor \emph{et al.}~\citep{ruttor2013approximate}, we employ a sparse kernel approximation for the drift $f$ by optimising the function values over a sparse set of $S$ inducing points $\{Z_i\}^{S}_{i=1}$.
We obtain the resulting drift from
\begin{equation}
    \hat{f}_S(x) = k^f(x,\mathcal{Z}) \left( I + \Lambda \, \mathcal{K}_S  \right)^{-1} \mathbf{d},
\end{equation}
where we have defined introduced the notation $\mathcal{K}_S \dot{=} k^f(\mathcal{Z},\mathcal{Z}) $
\begin{equation}
    \Lambda = \frac{1}{\sigma^2} \mathcal{K}^{-1}_S \left(   \int k^f(\mathcal{Z},x) A(x) k^f(x,\mathcal{Z}) \text{d}x \right)   \mathcal{K}^{-1}_S.
\end{equation}

\begin{equation}
    \mathbf{d} = \frac{1}{\sigma^2} \mathcal{K}^{-1}_S \left(   \int k^f(\mathcal{Z},x) B(x) \text{d}x \right)   \mathcal{K}^{-1}_S,
\end{equation}

We employ a sample based approximation of the densities in Eq.~\ref{appeq:drift_measure} resulting from the particle sampling of the path measure $Q$.
Thus by representing the densities by samples, we can rewrite the density $p_{t}(x)$ in terms of a sum of Dirac delta functions centered around the particles positions
$$p_{t}(x) \approx \frac{1}{N} \sum^N_{j=1} \delta(x - X_j(t)),$$ and replace the Riemannian integrals with summation over particles. Here $X_j(t)$ represents the position of the $j$-th particle at time point $t$.


 \section{Details on numerical experiments}\label{sec:details}
 We simulated a two dimensional Van der Pol oscillator
 with drift function 
 \begin{align}
        f_1(x,y) &= \mu(x - \frac{1}{3}x^3-y)\\
        f_2(x,y) &= \frac{1}{\mu}x,
 \end{align}
 starting from initial condition $ x0 = [1.81, -1.41]$
and under noise amplitudes $\sigma=\{0.25, 0.50, 0.75, 1.00\}$ for total duration of $T=\{500, 1000\}$ time units. The employed inter-observation intervals $\tau\in[80,   320] \times dt$. The last inter-observation interval exceeds the half period of the oscillator and thus samples only a single state per period. This resulted in erroneous estimates. 
In this setting this indicates the upper limit of $\tau$ for which we can provide estimates. However for any inference method, if the observation process samples only one observation per period, identifying the underlying force field without additional assumptions is not possible with temporal methods. The discretisation time-step used for simulation of the ground truth dynamics, and path augmentation $\delta t=0.01$.
For sampling the controlled bridges we employed $N=100$ particles evolving the associated ordinary differential equation as described in~\cite{maoutsa2022deterministic, maoutsa2021deterministica, maoutsa2020interacting}. The logarithmic gradient estimator used $M=40$ inducing points. The sparse Gaussian process for estimating the drift was based on a sparse kernel approximation of $S=300$ points. In the presented simulation we have employed a weighting parameter $\beta = 0.5$ (Eq.~\ref{apeq:free_energy2}). This provides a moderate pull towards the invariant density. The example in Figure~\ref{fig:path_augmentations} was constructed with $\beta=1$ and provides a better approximation of the transition density, than $\beta=0.5$.

For identifying geodesic curves in settings where only one observation is collected per oscillation period, we have to bias the method that computes the geodesics to identify the geodesic towards the correct direction of movement, that in this case is not always the shortest curve on the manifold between two consecutive observations. To that end we assigned to each observation a phase-like variable and considered for the computation of the geodesics only the observations that had properly valued phases given the direction of movement and the phases of the two end-point observations.
For example, for a counter-clockwise rotation and for phases at the end points of the bridge $\phi_k=\phi(\mathcal{O}_k)$ and $\phi_{k+1}=\phi(\mathcal{O}_{k+1})$, we construct the geodesics only on the observations that have phases within thin $[\phi_k, \phi_{k+1}]$ if $\phi_k< \phi_{k+1}$, or within $[\phi_k, 1] \cup [0, \phi_{k+1}]$ if $\phi_k> \phi_{k+1}$. With this approach we consider essentially only the relevant part of the manifold that aligns with the direction of movement. With the function $\phi(\cdot):\mathcal{R}^D \rightarrow [0,1]$ we denote the function that assigns a phase variable to each observation. Here for the Van der Pol oscillator we considered $\phi(\mathbf{x}) = \frac{\text{arctan2}(\frac{\mathbf{x}^{(2)}}{\mathbf{x}^{(1)}}) + \pi}{2 \pi} $. An alternative option for assigning phase to each observation is measuring the angle between the Hilbert transform of $\mathbf{x}^{(i)}$ and $\mathbf{x}^{(i)}$ itself, and rescaling the values to the $[0,1]$ interval. Here $\mathbf{x}^{(i)}$ denotes one dimensional component of the observations.

\section{Future directions}~\label{future}

The study of topological and geometric properties of invariant densities induced by stochastic dynamical systems is a relatively unexplored area with limited research available. Cong and Huang~\citep{cong1997topological} approached the study of topological properties of systems perturbed by noise through the lens of random dynamical systems, which, under certain conditions and assumptions, can be translated to the language of stochastic dynamical systems. This direction has great potential for future investigation, especially in conjunction with delay embedding approaches devised for random dynamical systems~\citep{stark2003delay} for estimation of the dimensionality of the invariant manifold.

One potential direction for future research is to directly employ Riemannian Langevin dynamics~\citep{wang2020wasserstein} to construct the augmented paths on the approximated metric. Alternatively, future directions for inference at the low sampling rate limit may tackle the problem through an operation learning perspective. Neglecting geometric considerations, constructing augmented paths for each inter-observation interval essentially solves the same control problem multiple times, only with different initial and terminal constraints. Thus neural network operator learning approaches like the neural operator used in~\cite{li2020fourier, li2020neurala} that can provide generalizations for different initial and possibly terminal conditions may be relevant for tackling the inference problem through path augmentation in a more computationally efficient way.

Further promising future directions may consider path augmentation in terms of Schr\"odinger bridges that would account for noisy observations (here we implicitly assume no noise in the observations or small Gaussian noise). ~\cite{tamir2023transport} propose an efficient algorithm for introducing path constraints in the Schr\"odinger bridge problem that may be employed in our framework to account for geometric inductive biases for inference of stochastic dynamics.

\section{Additional figures}

\begin{figure}[h!]
\centering
\includegraphics[width=0.68\textwidth]{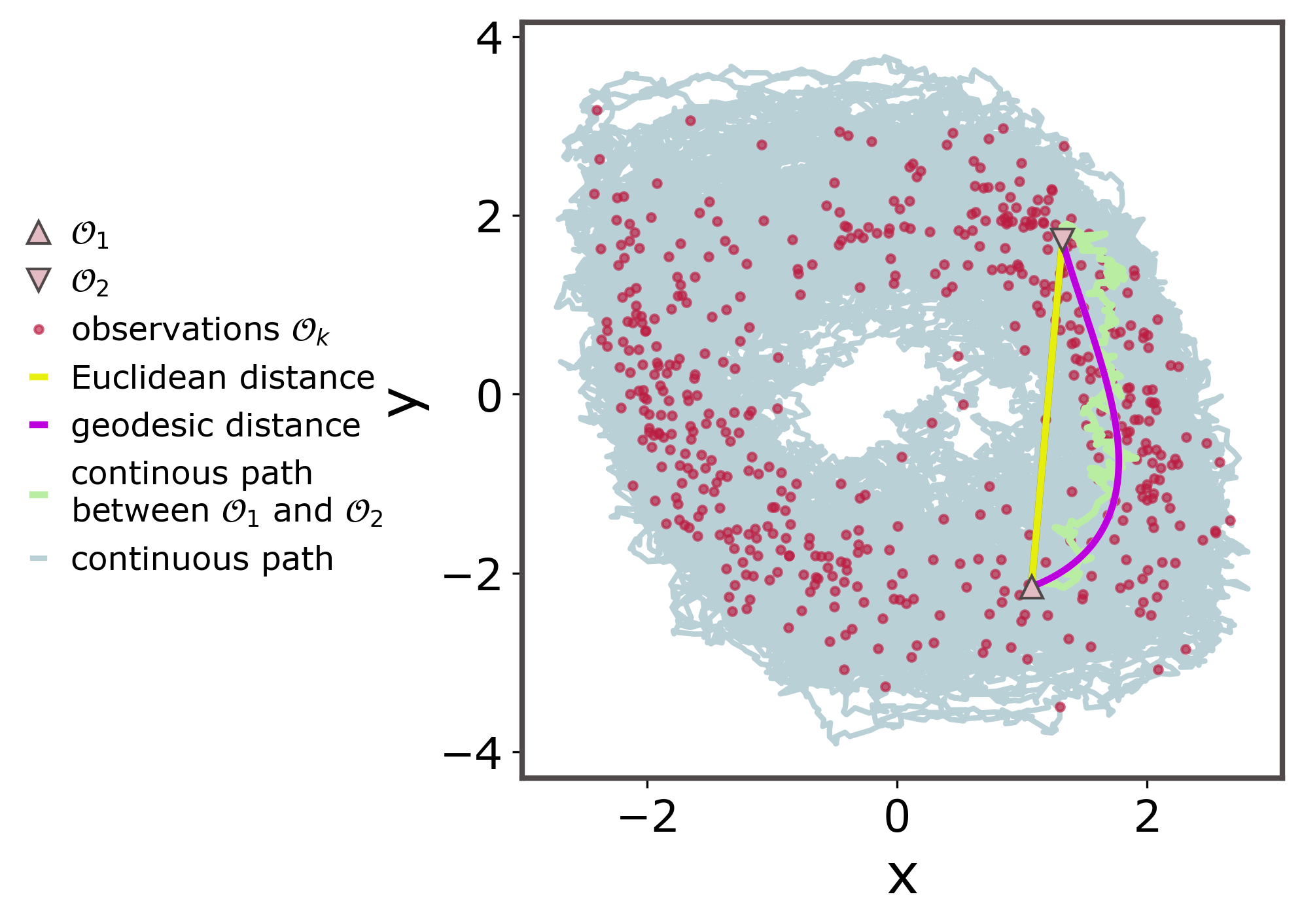}
  \caption{ \textbf{Considered state increments for low frequency observations under Gaussian likelihood assumptions.}  Euclidean distance (\emph{yellow line}) - used to compute the state increments between successive observations - does not account for the curvature of the invariant density. The geodesic curve (\emph{purple line}) provides a better approximation of the unobserved state of the system between successive observations (\emph{light green line}).}
  \label{fig:distances}
\end{figure}


\end{document}